\documentclass{article}
\usepackage{waspaa21,amsmath,graphicx,url,times}
\usepackage{color}
\usepackage{xcolor,comment}
\usepackage{subcaption}
\usepackage{multirow}
\usepackage{booktabs,tabularx}
\newcommand\note[1]{\textcolor{red}{#1}\PackageWarning{MyWarnings}{#1}}

\title{A Multi-Head Relevance Weighting   Framework for \\ Learning Raw Waveform Audio Representations}

\name{Debottam Dutta$^{*}$,
      Purvi Agrawal$^\$$, 
      Sriram Ganapathy$^{*}$, 
      }
\address{$^*$Learning and Extraction of Acoustic Patterns (LEAP) Lab, Indian Institute of Science, Bangalore, 560012 \\ 
         $^\$$Microsoft India Development Center, Hyderabad, India \\ 
}

\begin{document}

\ninept
\maketitle

\begin{sloppy}

\begin{abstract}
In this work, we propose a multi-head relevance weighting framework to learn audio representations from raw waveforms. The audio waveform, split into windows of short-duration, are processed with a 1-D convolutional layer of cosine modulated Gaussian filters acting as a learnable filterbank. The key novelty of the proposed framework is the introduction of multi-head relevance on the learnt filterbank representations.   Each head of the relevance network is modelled as a separate sub-network.  These heads perform representation enhancement by generating weight masks for different parts of the time-frequency representation learnt by the parametric acoustic filterbank layer.   The   relevance weighted representations are   fed to a neural classifier and the whole system is trained jointly for the audio  classification objective. Experiments are performed on the  DCASE2020 Task 1A  challenge as well as the Urban Sound Classification (USC) tasks. In these experiments, the proposed approach yields relative improvements of  $10$\% and $23$\% respectively for the DCASE2020 and USC datasets over the mel-spectrogram baseline. Also, the analysis of multi-head relevance weights provides insights on the learned representations.
\end{abstract}


\section{Introduction}
\label{sec:intro}
Recent advances in deep learning have shown substantial improvements in various domains including vision, language and speech~\cite{lecun2015deep}. The study of deep networks to learn meaningful representations from data, termed as deep representation learning,   has received wide interest in the last few years~\cite{bengio2013representation}. Specifically natural language~\cite{mikolov2013efficient} and computer vision tasks~\cite{chen2020simple} have shown promising methods for representation learning. While similar efforts have also been attempted for speech and audio signals~\cite{sainath}, knowledge driven representations, like the    mel-spectrogram, continue to be the most dominant approach in speech and audio related tasks. 

The conventional hand engineered speech features were inspired by multiple psycho-acoustic experiments~\cite{stevens}. Because of the consistent performances shown by mel-filterbank features in speech and speaker recognition  tasks, they also find applications in many of the audio-classification tasks. For representation learning from raw waveforms, early works in \cite{tuske2014acoustic}, \cite{sainath} used raw waveforms and power spectra respectively to learn filterbank parameters. Recently, unsupervised learning of filters were also explored by Agrawal et. al. \cite{Agrawal2019,agrawal2017unsupervised}. Schneider et. al. \cite{schneider} made use of dilated convolutions to learn pre-trained representations from raw audio in a self-supervised framework. Further, many speech related tasks such as speech separation \cite{mesgarani}, have successfully adopted learnable front-ends. Further, learnable front-ends have been explored recently by Zeghidour et. al.~\cite{zeghidour2021leaf}. 

In this work, we propose a multi-head relevance weighting based framework to learn audio representations directly from the raw waveform. This work advances the previous work on speech representation learning ~\cite{agrawal2020interpretable,agrawal2019unsupervised,agrawal2019modulation}.  The raw audio signal is windowed into short-time regions and a learnable 1-D convolutional layer is applied. The kernels of this layer are parameterized as cosine modulated Gaussian filters~\cite{agrawal2021thesis}. Following the 1-D convolution and average-pooling operation, a multi-head relevance weighting network is designed which generates self-attention mask for parts of the representation. The relevance weighted representations from the proposed framework (learned 2-D time-frequency representations) are used in the sound classification network. The entire network of filterbank parameters, relevance weights, and event classification network are learned jointly for the task. 

Experiments are performed on acoustic scene classification in DCASE2020 challenge Task 1A~\cite{DCASE2020_dataset} as well as the urbansound classification~\cite{usc_dataset} task. In these experiments, we show that proposed approach yields representations that improve the performance relatively by up to $23$ \% in terms of classification accuracy over the baseline log-mel filter representation. We also compare the proposed approach with SincNet representations~\cite{ravanelli2018speaker}. 

\section{Related Work}
Sainath et. al. \cite{sainath} proposed a framework to learn filters from the input power spectrum of the signal. In  the followup efforts, Hoshen et. al.  \cite{hoshen} investigated the learning of CNN filters directly from raw waveform which are initialized with Gammatone filters. In the direction of interpretable filterbank learning, Zeghidour et. al.  \cite{zeghidour} initialized filterbanks with Gabor wavelets. Recently, the SincNet approach proposed by Ravanelli et. al.  \cite{sincnet} considers the convolution kernels as sinc filters and learns only the low and high frequency cut-off. The proposed work in this paper is inspired by Agrawal et. al. \cite{agrawal2020interpretable}, where the approach explored  cosine modulated Gaussian filterbank. The Gaussian kernel has better time-frequency localization compared to sinc filters.  In this work, we also consider a more generalized relevance weighting approach through multi-head sub-networks.  
The motivation comes from prior works on Mixture of Experts (MoE) models \cite{jacobs1991adaptive,jordan1994hierarchical}. 

\section{FilterBank Learning with Multi-head Relevance weighting}
\label{sec:format}

The block schematic of the model is shown in Figure \ref{framework}. The proposed architecture is motivated by prior work on speech representation learning~\cite{agrawal2020interpretable,agrawal2021thesis}. 

\subsection{Acoustic filterbank layer}
For this layer, we consider the 1-D convolution layer   consisting of cosine-modulated Gaussian kernels. The input to this layer are the $T$ frames (windowed segments of audio) of raw audio files where each frame contains $S$ samples. This matrix of size $S \times T$ is passed through a 1-D convolutional layer where the kernels are parameterized by the cosine modulated Gaussian filter \cite{agrawal2020interpretable}, 
\begin{equation}
    g_i(n) = \cos{2\pi \mu_i n} \times \exp{(-n^2\mu_i^2/2)}, 
\end{equation} 
where $g_i(n)$ is the i-th kernel $(i = 1,2,..,F)$ and $\mu_i$ is the centre frequency of the \textit{i}-th kernel. The kernels perform convolution in each frame and generate $F$ feature maps. These feature maps are then squared, average pooled and log transformed. This way, from each frame, a $F$ dimensional feature is obtained. For the $T$  frames, the $F \times T$ shaped representation $\boldsymbol{x}$ is generated which we call the learned time-frequency (t-f) representation.  The learned t-f representation is also ordered in the increasing order of the center frequency $\mu _i$ before the relevance weighting. 


\begin{figure}[t!]
        \centering
	   \includegraphics[trim={0.3cm 0.1cm 0cm 0cm}, clip, width=0.48\textwidth]{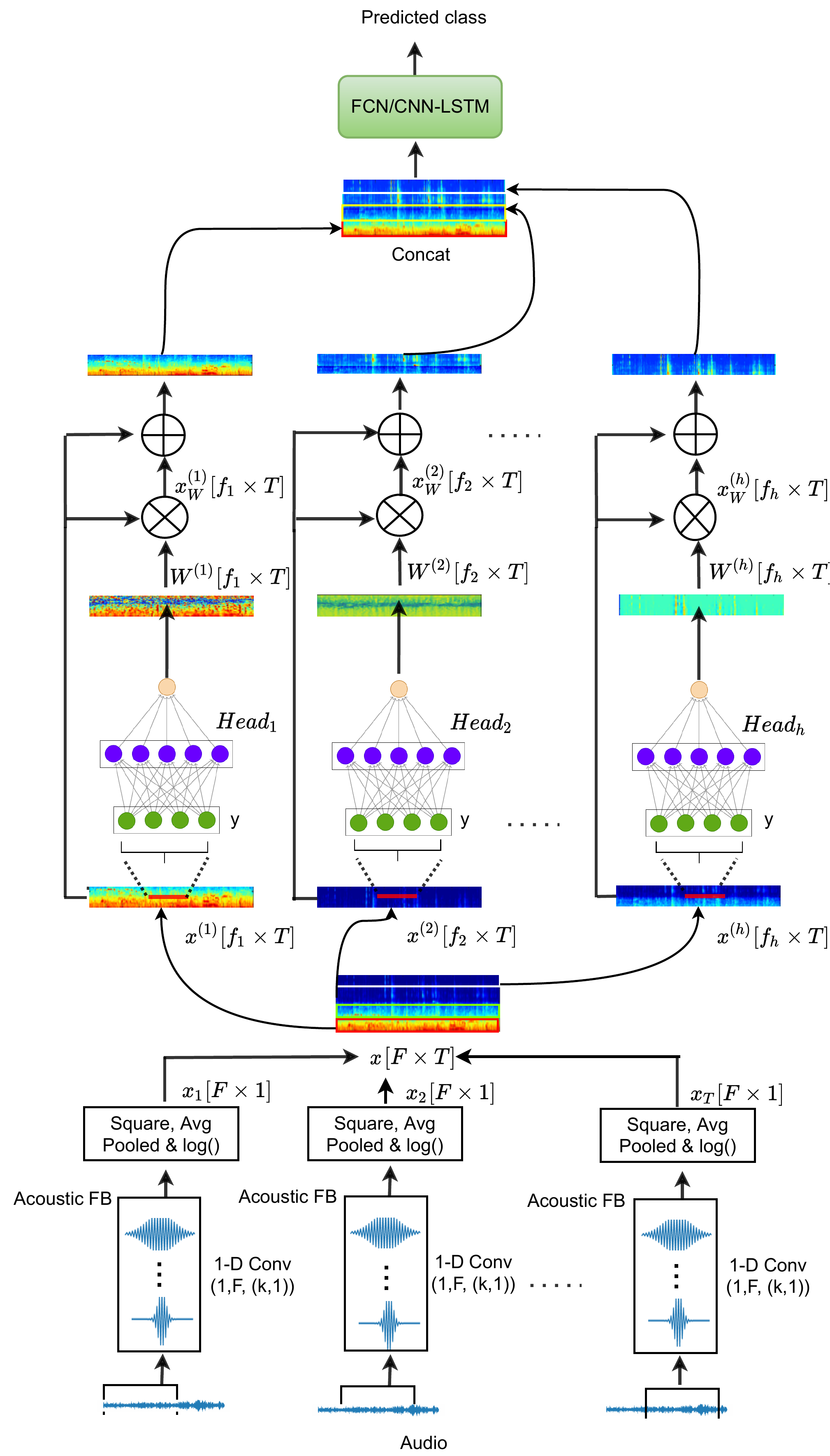}
	   \vspace{-0.5cm}
	    \caption{Proposed multi-head relevance weighting framework for audio representation learning. }
	    \label{framework}
\vspace{-0.5cm}
\end{figure}
\begin{table*}[t]
    \centering
    \caption{\textcolor{black}{Class-wise accuracy(\%) of mel, learned features over validation data. Differenet classes are Airport(AP), Bus(BS), Metro(MT), Metro station(MS), Urban park(PK), Public square(PS), Shopping mall(SM), Street pedestrian(SP), Street traffic(ST) and Tram (TM).}
    }
    \label{tab:dcase_results}
    \vspace{-0.15cm}
    \resizebox{0.78\textwidth}{!}{
   
    \begin{tabular}{c|c|c|c|c|c|c|c|c|c|c|c||c}
    \hline
    \multirow{2}{*}{Representations} & \multicolumn{12}{c}{{Accuracy} (\%)}  \\ \cline{2-13}
    & AP   & BS   & MT   & MS   & PK   & PS   & SM   & SP   & ST   & TM   & Avg & Avg. held-out  \\ \midrule
    Mel & \textbf{55.7} & 84.5 & 63.6 & 71.0 & 82.4 & 54.8 & \textbf{71.0} & 47.8 & 89.5 & 75.0 & 69.5 & 85.2\\
    SincNet~\cite{sincnet} & 46.9 & 83.8 & 70.3 & 71.0 & \textbf{85.5} & 55.5 & 65.3 & 49.1 & 85.8 & 70.9 & 68.5 & 86.5\\
    Cos-Gauss~\cite{Agrawal2019}    & 54.3 & 84.8 & \textbf{77.4} & 71.4 & 82.9 & \textbf{62.3} & 69.4 & 47.5 & 87.2 & 74.3 & 71.1 & 88.5 \\
    Cos-Gauss + 2head-rel. (Prop) & 53.0 & \textbf{86.1} & 76.8 & \textbf{72.0} & 84.8 & 58.9 & 69.7 & \textbf{56.6} & \textbf{90.6} & \textbf{76.7} & \textbf{72.6} & \textbf{89.6} \\
    
    \bottomrule
    \end{tabular}
    }
    \vspace{-0.1cm}
\end{table*} 
\begin{table*}[t]
    \centering
    \tiny
    \caption{\textcolor{black}{Classifier accuracy (\%) in UrbanSound8K database for different sound categories: Air conditioner (AI), Car horn (CA), Children playing (CH), Dog bark (DO), Drilling (DR), Engine idling (EN), Gun shot (GU), Jack hammer (JA), Siren (SI) and Street music (ST).}
    }
    \label{tab:usc_results}
    \vspace{-0.15cm}
    \resizebox{0.73\textwidth}{!}{
   
    \begin{tabular}{c|c|c|c|c|c|c|c|c|c|c|c}
    \hline
    \multirow{2}{*}{Representations} & \multicolumn{11}{c}{{Accuracy} (\%)}  \\ \cline{2-12}
    & AI & CA & CH & DO & DR & EN & GU & JA & SI & ST & Avg\\ \hline
    Mel   & 32 & 61 & 50 & 65 & 45 & 44 & 54 & 44 & 66 & 61 & 52\\
    Cos-Gauss \cite{agrawal2020interpretable} &  32 & \textbf{67} & 60 & 68 & 63 & 55 & \textbf{82} & 48 & \textbf{71} & 65 & 58\\
    Cos-Gauss + 2head-rel. (Prop)    & \textbf{37} & 37 & \textbf{62} & \textbf{69} & \textbf{70} & \textbf{64}	 & 79 & \textbf{71} & 60 & \textbf{68} & \textbf{63}\\ 
    \hline
    \end{tabular}
    }
    \vspace{-0.2cm}
\end{table*}
\subsection{Relevance weighting}
The multi-head relevance weighting method is proposed to enhance different parts of the learned t-f representation.  Each head of the multi-head relevance network is a $2$ layer fully connected network with sigmoidal activation in the hidden and output layers. Each head generates a soft mask for different parts of the learned spectrogram. We use a frequency based splitting of the t-f presentation to generate parts of the spectrogram used in the heads. For example, the model with $ h$ heads splits $\boldsymbol{x}$ along frequency axis into $h$ non-overlapping segments, $ \boldsymbol{x}^{(1)}, \boldsymbol{x}^{(2)},...,\boldsymbol{x}^{(h)}$.  Let, $f_{i}\times T$ be the shape of i-th segment, $\boldsymbol{x}^{(i)}$ which is fed as an input to i-th head. 

The relevance network for each head receives one sample from each time-frequency bin along with a temporal context of $(2c +1)$ from that sub-band to generate the relevance weight for that sample. Here, the context window size, $c$ is a hyper parameter in our experiments. The output of the relevance weighting network is, 
\begin{equation}
    W^{(i)}_{k,j} = \sigma(\Omega^{(i)}_{2}(\sigma(\Omega^{(i)}_{1}y_{k,j}+b^{(i)}_{1}) + b^{(i)}_{2}))
\end{equation}
where $ W^{(i)}_{k,j}$ is the entry in $k^{th}$ row and $j^{th}$ column of the weight mask $W^{(i)}$, $y_{k,j}$ is the $2c+1$ dimensional input vector $[\boldsymbol{x}^{(i)}_{k,(j-c)}, ..,\boldsymbol{x}^{(i)}_{k,j}, .., \boldsymbol{x}^{(i)}_{k,(j+c)}]^T $, $\Omega^{(i)}_1,b^{(i)}_1$ and $\Omega^{(i)}_2,b^{(i)}_2$ are weight matrices and biases for the first and second layer respectively. 
The same network is shared across all the elements of $\boldsymbol{x}^{(i)}$ to generate $W^{(i)}$ of the same shape $f_{i} \times T$. The weighted i-th  segment, $\boldsymbol{x}_W^{(i)}$ is obtained by element-wise multiplication of the weight mask $W^{(i)}$ with $\boldsymbol{x}^{(i)}$. At this stage, we call $\boldsymbol{x}_W^{(i)}$ the enhanced t-f representation segment for the i-th head. In total, the $h$ enhanced t-f segments will be generated from each of $h$ heads. Then,  skip-connection is used to add each part with the corresponding regions of the input representation, $\boldsymbol{x}^{(i)}$. Finally, all the representation segments are spliced to form the inputs to the audio classification network.

The frequency splitting method introduced in this section is similar to the generation of weights in a smaller dimensional representation space \cite{transformer} and late fusion of separate frequency paths \cite{McDonnel}. The frequency splitting method described here directly splits the time-frequency representation into multiple parts and uses a shared relevance sub-network for all the sub-bands within a split. 

The data augmentation methods such as SpecAug \cite{Park_2019}, random cropping, scaling etc. which are performed over spectrogram, are applied to individual heads separately. 
The augmented features are then fed to the rest of the classifier network for downstream task. We use number of Gaussian filters $F=80$ and kernel size $k=705$ ($16$ms for audio signal sampled at $44.1$kHz). The window length $S=2048$ ($46$ms)  and $S=1102$ ($25$ms) are used for DCASE2020 challenge and USC tasks respectively.   The rest of the classification network used is either a fully convolutional network (FCN)  for   the DCASE2020 dataset or a convolutional long short-term memory (CLSTM) based  network for the USC dataset. 

\section{Experiments and Results}

\subsection{Acoustic scene classification}
All the experiments for acoustic scene classification (ASC) task are performed on DCASE2020 Task 1A development dataset \cite{DCASE2020_dataset}. This dataset contains around $14000$ training samples and around $3000$ test samples which are recorded over $12$ European cities in $10$ different acoustic scenes. We use the officially provided train and validation sets for all the experiments and these sets are balanced across all the classes. The audio samples come from $3$ real devices A, B and C and $6$ simulated devices, s1-s6 of which s4-s6 are only part of the test set. A separate set of $6,000$ recordings are also used as a blind held-out set to compare the models (reported in last column of Table~\ref{tab:dcase_results}).  Along with the original train set, we perform different augmentations to increase the size of the dataset. The augmentation strategies  include (i) Spectrum correction \cite{speccorr} (ii) pitch shift, (iii) speed change, (iv) addition of random noise and (v) mixing of audio signals from the same class. All these strategies are performed at the audio level and they increase the training set size to around $80$K samples. Also, the SpecAug approach \cite{Park_2019}, random-cropping  \cite{McDonnel} and mix-up \cite{mixup} are applied on the spectrogram level to increase the robustness of the model learning. The spectrogram level augmentations do not increase the data set size and the random-crop is performed only along time axis. 

The baseline model used for the ASC task is the fully convolutional network (FCN)  classifier \cite{hu2020devicerobust}. This FCN model is a VGG-like \cite{Simonyan15} architecture with $9$ convolutional layers with small kernels (around $12$M parameters) as described in Hu et. al~\cite{hu2020devicerobust}. 
For the log-mel representations used in the baseline system, short time processing of audio signals are performed with window lengths of $2048$ and hop length of $1024$ samples. The windowed signal is passed through a mel filterbank to obtain $80$ log-mel energy features. These features are appended with delta, delta-delta features to obtain a final input of dimension  $80\times423\times3$. A scaling operation to convert the values to the range of $[0,1]$ is   performed before feeding   to FCN network. 

For the representation learning experiments, all the parameters are set according log-mel baseline for fair comparison. In our proposed multi-head network approach, we calculate delta-delta features after the skip-add connection from individual representation segments and then apply SpecAug and random-crop over them. All the experiment setups are implemented in PyTorch  \cite{pytorch}. 
The models are learned with stochastic gradient descent (SGD)  with momentum. A cosine annealing warm restart is used as learning rate scheduler to train all the models. The maximum and minimum learning rate used for the scheduler are $0.1$ and $10^{-5}$ respectively. 

Table \ref{tab:dcase_results} shows the performance of different models.The log-mel baseline model (Mel) provides $69.5$ \% accuracy on the test data. When the log-mel features are replaced by the learnt cosine modulated Gaussian features (Cos-Gauss), the accuracy is improved to $71.1$ \%. Further, the 2-head relevance network accuracy improves the performance over the baseline system with a $3.1$ \% absolute gain over the mel-baseline (relative improvement of $10$ \%). Further, on the separate held-out set (last column of Table~\ref{tab:dcase_results}), we observe that the proposed model gives consistent performance gains. 
\subsection{Urban sound classification}

The Urban Sound classification (USC) task~\cite{usc_dataset} contains  urban sound events. The UrbanSound8k dataset contains $8732$ sound clips sampled at $44.1$ kHz, with duration up to $4$s. For feature extraction, $25$ms window length and $10$ms hop length is used. Model evaluation is performed on the official predefined  $10$-fold cross validation setup. Averaging samples from all the folds we get $7.8$K samples for training and around $900$ samples for testing. We use the same multi-head relevance weighting setup with a different back-end classifier for the USC task. Here, the backend classifier is based on CNN-LSTM network having a stack of 1 convolutional layer (with $40$ filters and kernel size (5,5)), a $2$-layer LSTM network (with $256$ cells) followed by a fully connected layer of size $256$. 

Table \ref{tab:usc_results} reports the performance of the mel-baseline and the proposed multi-head approach.  For the proposed multi-head relevance network, we observe  a $5$ \% absolute improvement in average accuracy over the learned acoustic filterbank features~\cite{agrawal2020interpretable} and an absolute $11$ \% improvement over the baseline using mel-filterbank features. Also, as seen from the class-wise accuracy, the $2$-head relevance network is able to achieve improved accuracy in $7$ out of $10$ classes over the mel baseline system.  The car-horn sound had smaller number of training data (412 samples) compared to other dominant classes.  In the analysis, we found that the model was substituting this class with street-music class.
\begin{figure}[t!]
        \centering
	    \includegraphics[width=0.4\textwidth]{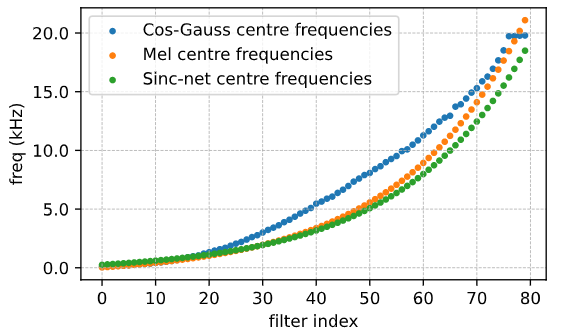}
	    \vspace{-0.2cm}
	    \caption{Distribution of the centre frequencies of mel filterbank (orange), learnt Cos-Gauss filterbank (blue) and learnt SincNet filterbank (green) for DCASE challenge task.}
	    \label{fig:center_freq}
\end{figure}

\begin{figure}[t!]
      
        \centering
	    \includegraphics[trim={1.1cm 2.1cm 0.9cm 0.8cm}, clip, width=\linewidth]{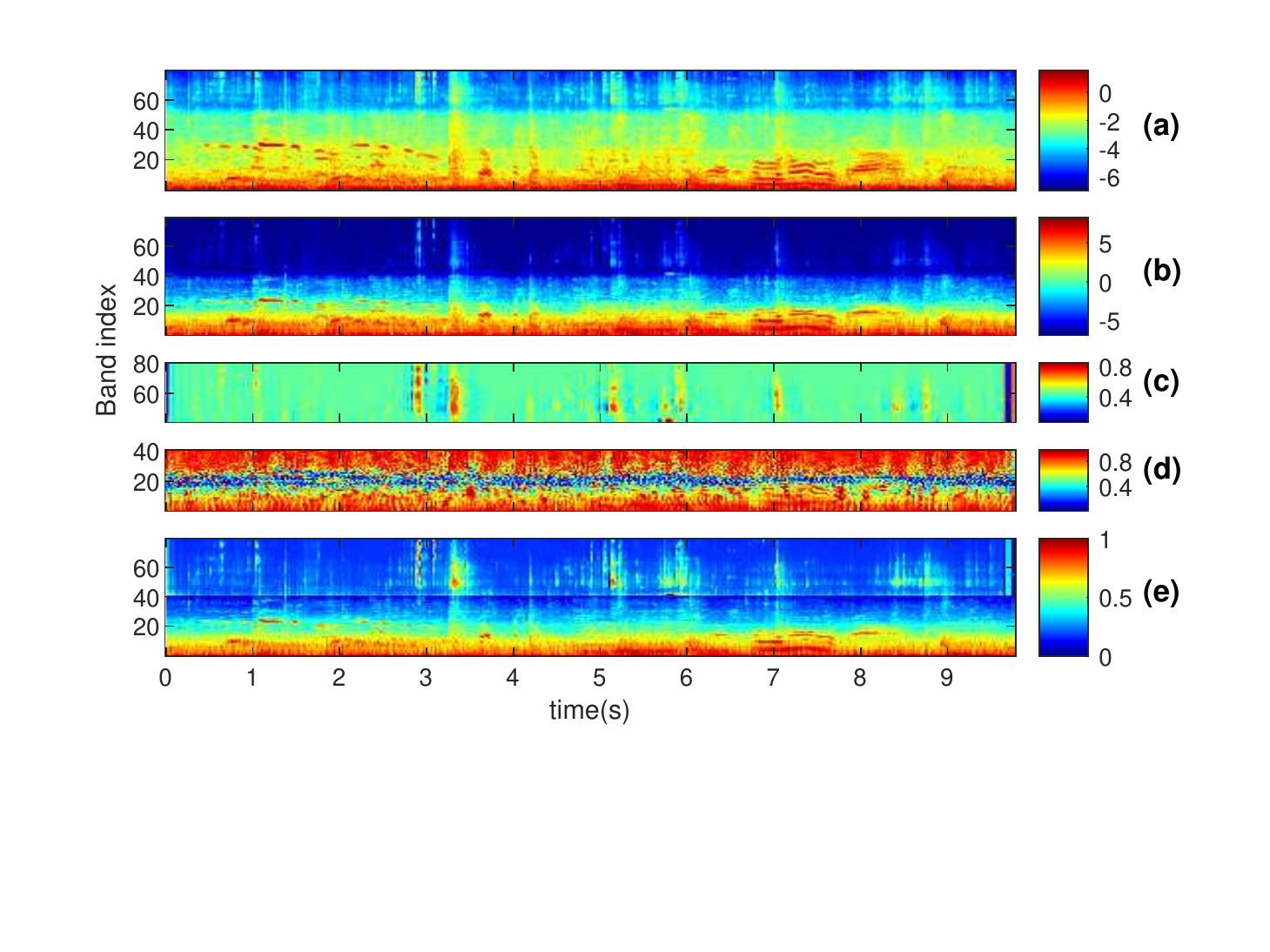}
	    \vspace{-0.7cm}
	    \caption{(a) Mel spectrogram for audio file \textit{airport-barcelona-0-0-a.wav} from DCASE dataset,
     (b) output of acoustic FB layer (learnt t-f reprentation $\boldsymbol{x}$ in Figure \ref{framework}),   
     (c) mask produced for the last $40$ frequency bands by $2^{nd}$ head $(\boldsymbol{W}^{(2)})$, 
     (d) mask produced for first $40$ frequency bands by $1^{st}$ head $(\boldsymbol{W}^{(1)})$, and (e) final representation with relevance weighting and scaling.}
	    \label{fig:specgram}
\vspace{-0.5cm}
\end{figure}
\subsection{Discussion}
\textbf{Center frequency profile} - Figure \ref{fig:center_freq} shows the distributions of centre frequencies of learnt Gaussian filter-bank, mel filter-bank and SincNet \cite{sincnet} for ASC task. The SincNet and mel filter center frequencies are similar except in very high frequencies where the SincNet approach prefers lower center frequencies than the mel-filter bank. The cosine modulated Gaussian filterbank  matches the mel curve at very low frequencies, but allocates more filters in high frequency regions. This behaviour is expected as acoustic scenes contain lot of high frequency sources such as traffic sounds, beat sounds and background music. 
The modified center frequency profile may also explain the performance improvements seen in Table~\ref{tab:dcase_results} (comparing rows 1,2 and 3), where the Cos-Gauss filters improve by an absolute margin of $2.6$ \% over the SincNet approach.  

\textbf{Time-frequency representations} -  Figure~\ref{fig:specgram} provides the time-frequency representations from the proposed approach. The proposed approach to filterbank learning emphasizes higher frequencies (more filters on the high frequency region). This attributes to broader blue upper region in Figure~\ref{fig:specgram}(b). 
The two relevance weights (masks) for the learned representation from a 2-head setting are shown in Figure~\ref{fig:specgram}(c) and (d). The two masks clearly show the time adaptivity of the weights. The audio file comes from an airport scene and around $3$,$5$,$6$ and $7$ sec regions, the audio contains sound of metal box closing and opening. The mask in Figure~\ref{fig:specgram} (c) asserts more weight on these acoustic events and suppresses the rest. 
The final weighted representations (Figure~\ref{fig:specgram} (d)) is seen to provide sharp time frequency localization of the events. 

\begin{table}[t!]
\caption{Analysis of DCASE model performance for different number of heads (*H), context window size $c$, aligned (A) and non-aligned (NA) feature in SpecAug (SA), type of relevance weighting architecture (R.Arch), different ways of network sharing for sub-band splitting (Split) and presence of skip-add connection (skpAdd). }
\label{tab:dcase_hyperPara_analysis}
\centering
\vspace{-0.1cm}
\resizebox{0.45\textwidth}{!}{
\begin{tabular}{ccccccc}
\toprule
Model                & $c$ & SA & R.Arch & Split. & skpAdd & Acc(\%) \\ \midrule 
\multirow{3}{*}{1H-L}  & 5    & - & FC     & -               & no & 71.4 \\ 
                      & 10    & - & FC     & -               & no & 71.5 \\ 
                      & 20    & - & FC     & -               & no & 70.7 \\ \midrule
\multirow{5}{*}{2H-L} & 10    & NA  & FC    & 40-40          & no & 71.2 \\                             & 10   & A  & FC    & 40-40     & no   & 72.1 \\
                      & 10    & A   & FC    & even-odd    & no   & 69.9 \\
                      & 10    & A   & FC    & 40-40    & yes    & \textbf{72.6} \\
                      & 10    & A   & Conv   & 40-40     & no     & 72.0 \\ \midrule
                      
3H-L                  & 10     &A    & FC    & 26-27-27     & no       & 69.6 \\ \midrule
4H-L                  & 10     &A    & FC    & 20-20-20-20    & no       & 69.3 \\ \midrule
80H-L                 & 10     &A    & FC    & No sharing    & yes      & 69.5 \\ 
\bottomrule
\end{tabular}
}
\vspace{-0.3cm}
\end{table}
\textbf{Hyper-parameter selection} - We experiment with various  design aspects and hyper-parameters of our model. Starting with the one relevance head (1H-L) model, from Table \ref{tab:dcase_hyperPara_analysis} it can be seen that context window length $c=10$ (context length used in relevance weight network) offers the best accuracy. 
When the number of heads is increased, we see a decrease in accuracy if the data augmentation (SpecAug  and random-crop) from the heads are not aligned (NA), i.e., when the time and frequency mask positions and cropping of time frames in random-crop are not aligned amongst the heads. Hence, with the aligned version of SpecAug and random-crop, the model gains by an absolute margin of $0.9$ \% in accuracy. We also experiment with the even-odd splitting (odd frequency bands as input to first head and even frequency bands as input to second head in the relevance network). These experiments show that relevance head splitting using 40-40, where the spectrum of first $40$ bands is used in the first relevance head and the next $40$ bands used in the second head, provides the best accuracy. We also considered using a 1-D convolution (Conv) layer (with $8$ kernels of size $1\times 3$) for the first layer of relevance head sub-network. Finally, results from $3$-head and $4$-head experiments are shown with roughly equal sub-band splits. A separate relevance head for all the $80$ sub-bands is also shown in the Table~\ref{tab:dcase_hyperPara_analysis}. Here, the relevance network for each sub-band is not shared with other sub-bands. The best performance is achieved with $2$-head relevance weighting and skip-add.

\textbf{Model complexity analysis} - 
For DCASE2020 Task1a dataset, the FCN classifier has around $11.78$M parameters. Our proposed front-end with learnable acoustic filterbank and a single relevance sub-network adds only $1.1$k parameters to the model (Cosine-Gaussian filterbank has only 80 mean trainable parameters). Thus in terms of total parameters involved, the best performing 2-head relevance net approach adds only $0.02$ \% extra parameters over the FCN baseline classifier. For the Urban sound classification task, the  total number of parameters in mel baseline is  $1.18$M while in the $2$-head proposed approach it increases to $1.19$M. This leads to a negligible increase of $0.06$ \% in the number of parameters. 
\section{Summary}
In this paper, we explore   audio representation learning   with an acoustic filterbank learning and multi-head relevance weighting. We introduce a novel  frequency splitting method to learn and enhance different parts of the time-frequency representation.  The proposed multi-head framework generates representations that emphasizes high frequency regions of the audio signal. The multi-head framework is shown to provide significant improvements in both DCASE2020 Task 1A and UrbanSound8k datasets.  
\bibliographystyle{IEEEtran}
\bibliography{refs21}
%
%
%
%
%
%
%
%
%

\end{sloppy}
\end{document}